\documentclass[10pt,english,aps,prd,twocolumn,superscriptaddress,nofootinbib,preprintnumbers,floatfix]{revtex4-2}
\usepackage[utf8]{inputenc}
\usepackage{amsmath}
\usepackage{amsfonts}
\usepackage{amssymb}
\usepackage{graphicx}
\usepackage{array}
\usepackage{booktabs}
\usepackage{multirow}
\usepackage{mathtools}

\usepackage{verbatim}
\usepackage[dvipsnames]{xcolor}
\usepackage{hyperref}   
\hypersetup{colorlinks=true, 
	citecolor=MidnightBlue, linkcolor=MidnightBlue, urlcolor=Cyan,
	pdftitle={Exact Ultrastatic Multi-Throat Wormholes: Geometry, Energy Conditions, and Optical Cavities},
	pdfauthor={Gabriel I. R\'ois, Francisco S. N. Lobo, and Manuel E. Rodrigues}}
\usepackage{tikz}
\definecolor{purple}{rgb}{1,0,1}
\definecolor{lime}{HTML}{A6CE39} 

\newcommand{\orcidicon}{%
	\begin{tikzpicture}
		\draw[lime, fill=lime] (0,0) 
		circle [radius=0.16] 
		node[white] {{\sffamily\tiny ID}};
		\draw[white, fill=white] (-0.0625,0.095) 
		circle [radius=0.007];
	\end{tikzpicture}	\hspace{-2mm}
}
\newcommand\orcidFrancisco{{\href{https://orcid.org/0000-0002-9388-8373}{\orcidicon}}}
\newcommand\orcidManuel{{\href{https://orcid.org/0000-0001-8586-0285}{\orcidicon}}}

\newcommand\orcidGabriel{{\href{https://orcid.org/0009-0007-3770-8976}{\orcidicon}}}

\begin{document}

\title{Exact ultrastatic multi-throat wormholes:\texorpdfstring{\\}{ }
Geometry, energy conditions, and optical cavities}

\author{Francisco S. N. Lobo\orcidFrancisco\!\!} 
\email{fslobo@ciencias.ulisboa.pt}
\affiliation{Instituto de Astrof\'{i}sica e Ci\^{e}ncias do Espa\c{c}o, Faculdade de Ci\^{e}ncias da Universidade de Lisboa, Edifício C8, Campo Grande, P-1749-016 Lisbon, Portugal}
\affiliation{Departamento de F\'{i}sica, Faculdade de Ci\^{e}ncias da Universidade de Lisboa, Edif\'{i}cio C8, Campo Grande, P-1749-016 Lisbon, Portugal}

\author{Manuel E. Rodrigues\orcidManuel\!\!}
\email{esialg@gmail.com}
\affiliation{Faculdade de F\'{\i}sica, Programa de P\'{o}s-Gradua\c{c}\~{a}o em 
	F\'isica, Universidade Federal do 
	Par\'{a},  66075-110, Bel\'{e}m, Par\'{a}, Brazil}
\affiliation{Faculdade de Ci\^{e}ncias Exatas e Tecnologia, 
	Universidade Federal do Par\'{a}\\
	Campus Universit\'{a}rio de Abaetetuba, 68440-000, Abaetetuba, Par\'{a}, 
	Brazil}

\author{Gabriel I. R\'{o}is\orcidGabriel\!\!}
\email{fc54507@alunos.ciencias.ulisboa.pt}
\affiliation{Instituto de Astrof\'{i}sica e Ci\^{e}ncias do Espa\c{c}o, Faculdade de Ci\^{e}ncias da Universidade de Lisboa, Edifício C8, Campo Grande, P-1749-016 Lisbon, Portugal}
\affiliation{Departamento de F\'{i}sica, Faculdade de Ci\^{e}ncias da Universidade de Lisboa, Edif\'{i}cio C8, Campo Grande, P-1749-016 Lisbon, Portugal}

\date{\today}

\begin{abstract}
We construct an exactly solvable hierarchy of static, spherically symmetric, wormholes with arbitrary finite throat multiplicity.  A
factorized prescription for the derivative of the areal radius places every
stationary sphere at a prescribed proper distance while retaining elementary
integrability and asymptotic flatness.  The order-$N$ geometry contains
exactly $N+1$ throats and $N$ anti-throats, arranged as an alternating chain
of necks and finite interior cavities.  Independently of this particular
hierarchy, the Einstein equations imply three universal ultrastatic results.
Every nondegenerate throat violates the radial null energy condition, whereas
every nondegenerate anti-throat satisfies the null, weak, dominant, and strong
energy conditions, with the strong-energy sum condition saturated.  The same
local combination $\rho+p_\ell$ fixes the radial stability of the associated
circular null orbit through
$V_{\rm null}''=8\pi L^2(\rho+p_\ell)/\mathcal R^2$; throats are therefore
unstable light rings and anti-throats are stable ones.  Finally, every finite
cell between consecutive stationary spheres, each exterior cell, and the
complete two-ended radial null geodesic carry strictly negative integrated
radial null energy.  The resulting spacetimes are exact multi-cavity optical
chains whose cavity centers are locally nonexotic even though no cell can
eliminate averaged radial exoticity.  The first nontrivial member
interpolates continuously from the Ellis--Bronnikov wormhole through a
quartically degenerate throat to a double-throat geometry and admits closed
expressions for its source, null average, branchwise Morris--Thorne form,
embedding criterion, and traversability scales.  Its vanishing
Arnowitt--Deser--Misner mass follows from the specific radial falloff of the
hierarchy, rather than from ultrastaticity alone.
\end{abstract}

\maketitle

\section{Introduction}
\label{sec:intro}

Traversable Lorentzian wormholes turn the geometry of a spatial bridge into
a sharp question about matter, causality, and wave propagation.  In the
canonical Morris--Thorne picture, the bridge is identified by a minimum-area
two-sphere joining two asymptotic regions \cite{MorrisThorne1988}.  The
possibility of converting a traversable bridge into a time machine also makes
the relation between wormholes, energy-condition violation, and chronology a
fundamental physical issue \cite{MorrisThorneYurtsever1988}.  The
invariant minimal-area and flare-out formulation was subsequently clarified
by Hochberg and Visser \cite{HochbergVisser1997}.  If $\ell$ is proper radial
distance and $\mathcal R(\ell)$ is the areal radius, a nondegenerate throat is
simply characterized by $\mathcal R'=0$ and $\mathcal R''>0$.  This global
proper-distance description becomes indispensable when $\mathcal R$ is not
one-to-one, because a single areal-radius chart then fails precisely where
the geometry becomes most interesting
\cite{Visser1989,Visser1995,Lobo2005,Lobo2007}.

Nothing in the global topology $\mathbb R\times S^2$ requires the area
profile to possess only one minimum.  Several necks may occur, separated by
local maxima of the area -- equators, bellies, or, as we shall call them,
\emph{anti-throats}.  Such structures arise in configurations involving
ghost scalars and neutron matter, Yang--Mills hair, bosonic matter, and
rotating matter
\cite{Dzhunushaliev2014NS,Hauser2014,Dzhunushaliev2014Boson,Hoffmann2018},
as well as in higher-curvature gravity \cite{Antoniou2020,Ibadov2020}.
Multiple throats and anti-throats have also been developed in black-bounce
geometries \cite{RodriguesSilva2023,RodriguesSilva2025} and reconstructed
from a phantom scalar coupled to nonlinear electrodynamics (NED)
\cite{Crispim2025}.  These examples show that a multi-throat profile is not a
geometrical curiosity.  In the ultrastatic subclass considered here, every
additional maximum of the areal radius is a stable circular null orbit and
the neighboring minima form optical barriers.  Beyond ultrastaticity the
null potential also depends on the redshift function, so extrema of
$\mathcal R$ alone need not be light rings.

Recent analytic constructions sharpen this point.  Suitable embedding data
can generate traversable wormholes systematically \cite{Rueda2022}; a
quartic embedding relation yields a two-throat, asymptotically flat geometry
with a branchwise shape function \cite{Cataldo2026}.  Localized deformations
of the Ellis--Bronnikov geometry can support a nonexotic central belly in a
string-cloud background \cite{Amaral2026}, while finite-mass two-throat
families display a rich global light-ring structure, strong lensing, and
resonant scalar transmission \cite{Pradhan2026}.  The relevant challenge is
therefore no longer to exhibit one further double-throat example.  It is to
obtain a hierarchy in which an arbitrary number of stationary spheres can be
placed, classified, and analyzed without surrendering exact control.

Circular null orbits of wormholes have a broader geometrical context.
General existence results show that traversable wormholes carry light rings
under mild stationarity and asymptotic assumptions \cite{Xavier2024}, while
reflection-symmetric spherical throats require a careful treatment of stable,
unstable, and degenerate circular orbits \cite{Tsukamoto2024}.  Multiple
photon and antiphoton spheres can produce a distinctive strong-lensing image
structure \cite{Shaikh2019}.  The present work complements these global
results: within the ultrastatic spherical sector it gives an exact local
identity between light-ring curvature and the radial null-energy combination,
and supplies a hierarchy in which every circular null orbit is known
algebraically.

This requirement motivates working with $\mathcal R'(\ell)$ rather than
guessing $\mathcal R(\ell)$ itself.  The zeros of the derivative are the
stationary spheres, and their multiplicities determine whether the extrema
are genuine or degenerate.  Factoring the derivative therefore turns the
desired geometry into algebraic input.  The remaining freedom must then be
used to enforce the asymptotic slopes $\mathcal R'\to\pm1$, regularity on the
entire real line, reflection symmetry, and elementary integrability.  These
requirements single out a particularly economical algebraic family: its
ordered polynomial roots prescribe all proper-distance locations, while its
denominator supplies exactly the degree needed for asymptotic flatness and
converts the integral into a finite sum of half-integer powers.  The ansatz is
thus not an arbitrary decorative choice; it is a constructive solution to
the inverse problem of designing a smooth, analytically tractable
multi-throat geometry.

An additional advantage of this approach is that it cleanly separates what
belongs to the chosen hierarchy from what follows from ultrastatic Einstein
geometry alone.  The sign of $\mathcal R''$ simultaneously classifies a
stationary sphere, fixes the radial null-energy combination, and determines
the radial stability of its circular null orbit.  Integrated identities then
show how local satisfaction of all pointwise energy conditions at an
anti-throat coexists with unavoidable averaged exoticity in every adjoining
cell.  The geometry consequently becomes an exactly controlled chain of
optical cavities rather than merely a profile with several extrema.
This local--global distinction is important: pointwise restoration of the
energy conditions at a cavity center does not remove the averaged null-energy
violation required along a complete traversing null ray
\cite{VisserKarDadhich2003,KarDadhichVisser2004}.
Appendix~\ref{app:comparison} compares the present construction with representative recent multi-throat models, clarifying its scope and distinguishing established features from the results developed here.

This work is organized as follows. In Sec.~\ref{sec:universal}, we derive the universal geometric and matter identities governing ultrastatic, spherically symmetric spacetimes. In Sec.~\ref{sec:hierarchy}, we introduce the factorized prescription for \(\mathcal{R}'_{N}(\ell)\) and construct an analytically tractable hierarchy with arbitrary finite throat multiplicity. Section~\ref{sec:N1} examines the \(N=1\) double-throat geometry in detail, including its energy conditions, averaged null-energy violation, optical structure, and embedding properties. In Sec.~\ref{sec:multicavity}, we extend the optical analysis to the full hierarchy and discuss the resulting multi-cavity null dynamics. The traversability conditions, physical interpretation, and domain of applicability of the solutions are addressed in Sec.~\ref{sec:scope}. Finally, Sec.~\ref{sec:conclusions} summarizes the principal results, discusses their broader implications, and outlines directions for future work. Appendix~\ref{app:source} reconstructs an Einstein--Maxwell--scalar realization of the same effective source, linking the geometrical identities to explicit field couplings without assuming a unique matter model. Appendix~\ref{app:comparison} collects the comparison with related multi-throat constructions.
Throughout we use geometrized units $G=c=1$, metric signature
$(-,+,+,+)$, and the Einstein equations $G_{\mu\nu}=8\pi T_{\mu\nu}$ with
vanishing cosmological constant.  Greek indices label spacetime components,
hatted indices refer to an orthonormal frame, and a prime denotes
$d/d\ell$ unless another derivative is displayed explicitly.

\section{Universal ultrastatic identities}
\label{sec:universal}

We begin with results that hold for any smooth positive areal radius
$\mathcal R(\ell)$ in the ultrastatic spherical metric
\begin{equation}
 ds^2=-dt^2+d\ell^2+\mathcal R^2(\ell)d\Omega^2,
 \qquad -\infty<\ell<+\infty.
 \label{eq:generalmetric}
\end{equation}
Here $t$ is the proper time of static observers, $\ell$ is the global proper
radial distance, $\mathcal R>0$ is the areal radius, and
$d\Omega^2=d\theta^2+\sin^2\!\theta\,d\phi^2$ is the line element on the unit
two-sphere.  A symmetry sphere at $\ell$ has area
$A(\ell)=4\pi\mathcal R^2(\ell)$.
The proper-distance chart is global as long as $\mathcal R>0$ and smooth.
A stationary area sphere satisfies $\mathcal R'=0$.  It is a throat if
$\mathcal R''>0$, an anti-throat if $\mathcal R''<0$, and degenerate if the
second derivative vanishes.

Regularity is transparent in this gauge.  The Kretschmann scalar is
\begin{equation}
 K\equiv R_{\mu\nu\rho\sigma}R^{\mu\nu\rho\sigma}
 =4\left[
 2\left(\frac{\mathcal R''}{\mathcal R}\right)^2
 +\left(\frac{1-\mathcal R'^2}{\mathcal R^2}\right)^2
 \right].
 \label{eq:generalKretschmann}
\end{equation}
Thus a smooth positive $\mathcal R$ with finite first two derivatives gives a
curvature-regular spacetime.  This observation will also make the global
regularity of the hierarchy immediate.

\subsection{Alternation and counting of stationary spheres}
\label{subsec:countinggeneral}

For a two-ended asymptotically flat geometry one has
\begin{equation}
	\begin{aligned}
		\mathcal R'(\ell)&\longrightarrow -1,
		&&\ell\to-\infty,
		\\
		\mathcal R'(\ell)&\longrightarrow +1,
		&&\ell\to+\infty.
	\end{aligned}
	\label{eq:Rprimeasymgeneral}
\end{equation}
Assume that the number of stationary spheres is finite and that every one is
nondegenerate.  Since a simple zero of $\mathcal R'$ changes its sign, the
first stationary sphere encountered from the left must change
$\mathcal R'$ from negative to positive and is therefore a throat.  The next
must be an anti-throat, and the extrema continue to alternate.  Because the
final asymptotic sign is positive, the last stationary sphere is again a
throat.  Thus
\begin{equation}
 N_{\rm T}=N_{\rm A}+1,
 \label{eq:generalcount}
\end{equation}
where $N_{\rm T}$ and $N_{\rm A}$ are the numbers of nondegenerate throats and
anti-throats, respectively.

Equation~\eqref{eq:generalcount} is purely geometrical.  In particular, a
smooth finite multi-throat wormhole cannot contain several throats separated
only by one central anti-throat.  Additional throats necessarily require
additional maxima.  Equivalently, every finite interval between consecutive
throats contains exactly one anti-throat.  This motivates the interpretation
of a multi-throat geometry as a chain of finite interior cavities or
``islands'' separated from one another and from the asymptotic regions by
narrower necks.

\subsection{Einstein equations and the stationary-sphere energy theorem}
\label{subsec:energytheorem}

Let
\begin{equation}
 T_{\hat\mu\hat\nu}=\operatorname{diag}(\rho,p_\ell,p_t,p_t)
 \label{eq:generalT}
\end{equation}
be the stress-energy tensor in the static orthonormal frame, where $\rho$ is
the energy density, $p_\ell$ the radial pressure, and $p_t$ the two equal
tangential pressures.  For this diagonal anisotropic source, the null energy
condition (NEC) is $\rho+p_i\geq0$; the weak energy condition (WEC) adds
$\rho\geq0$; the dominant energy condition (DEC) requires
$\rho\geq0$ and $\rho\geq|p_i|$; and the strong energy condition (SEC)
requires $\rho+p_i\geq0$ together with
$\rho+p_\ell+2p_t\geq0$, where $i\in\{\ell,t\}$.  The Einstein equations
give
\begin{align}
 8\pi\rho
 &=\frac{1-\mathcal R'^2-2\mathcal R\mathcal R''}{\mathcal R^2},
 \label{eq:generalrho}
 \\
 8\pi p_\ell
 &=\frac{\mathcal R'^2-1}{\mathcal R^2},
 \label{eq:generalpl}
 \\
 8\pi p_t
 &=\frac{\mathcal R''}{\mathcal R}.
 \label{eq:generalpt}
\end{align}
A first universal identity is
\begin{equation}
 \rho+p_\ell+2p_t=0,
 \label{eq:generalactivezero}
\end{equation}
which is the vanishing Tolman--Komar active-density combination associated
with the unit lapse.  In particular, Eq.~\eqref{eq:generalactivezero}
saturates the SEC sum condition everywhere, but it does not by itself fix the
other NEC inequalities.

The radial null combination is especially simple:
\begin{equation}
 \rho+p_\ell
 =-\frac{1}{4\pi}\frac{\mathcal R''}{\mathcal R}.
 \label{eq:generalradialNEC}
\end{equation}
Hence every nondegenerate throat violates the radial NEC and every
nondegenerate anti-throat satisfies it.  More can be said at an anti-throat.
At any stationary sphere $\mathcal R'=0$,
\begin{align}
 8\pi\rho_*&=\frac{1}{\mathcal R_*^2}-2\frac{\mathcal R_*''}{\mathcal R_*},
 \label{eq:stationaryrho}
 \\
 8\pi p_{\ell *}&=-\frac{1}{\mathcal R_*^2},
 \label{eq:stationarypl}
 \\
 8\pi p_{t*}&=\frac{\mathcal R_*''}{\mathcal R_*}.
 \label{eq:stationarypt}
\end{align}
If the surface is an anti-throat, $\mathcal R_*''<0$, and therefore
\begin{align}
 \rho_*&>0,
 \\
 \rho_*+p_{\ell *}&>0,
 \qquad
 \rho_*+p_{t*}>0,
 \\
 \rho_*-|p_{\ell *}|&>0,
 \qquad
 \rho_*-|p_{t*}|>0.
 \label{eq:antithroatECall}
\end{align}
Together with Eq.~\eqref{eq:generalactivezero}, this proves the following
statement:

\medskip
\noindent\textbf{Ultrastatic anti-throat energy theorem.}---
\emph{In four-dimensional Einstein gravity, every nondegenerate anti-throat
of a spherically symmetric, ultrastatic spacetime satisfies the NEC,
WEC, DEC, and SEC.  The SEC sum condition
$\rho+p_\ell+2p_t\geq0$ is saturated.}
\medskip

Thus the locally nonexotic central belly found in particular double-throat
models is not an accidental property of a special source: within the
ultrastatic class it follows directly from the sign of $\mathcal R''$.
Conversely, at every throat Eq.~\eqref{eq:generalradialNEC} is negative,
although the sign of $\rho$ itself remains model dependent.

\subsection{Exact energy-condition/light-ring correspondence}
\label{subsec:opticaltheorem}

By spherical symmetry a null geodesic may be placed in the equatorial plane
$\theta=\pi/2$.  Let $\lambda$ be an affine parameter and define
$\dot{\ }\equiv d/d\lambda$.  The Killing energy $E>0$ and azimuthal angular
momentum $L$ are
\begin{equation}
 E=\dot t,
 \qquad
 L=\mathcal R^2\dot\phi,
 \label{eq:generalEL}
\end{equation}
and the radial equation, with effective null potential $V_{\rm null}$, is
\begin{equation}
 \dot\ell^2+V_{\rm null}(\ell)=E^2,
 \qquad
 V_{\rm null}(\ell)=\frac{L^2}{\mathcal R^2(\ell)}.
 \label{eq:generalVnull}
\end{equation}
We use \emph{light ring} to mean a sphere foliated by circular null
geodesics.  Every stationary area sphere is therefore a light ring in the
ultrastatic geometry.  This conclusion is special to a constant lapse: for
$g_{tt}=-e^{2\Phi(\ell)}$, the potential becomes
$L^2e^{2\Phi}/\mathcal R^2$ and depends on both $\Phi$ and $\mathcal R$.
At a stationary sphere of the present geometry,
\begin{equation}
 V_{\rm null}''\big|_*
 =-\frac{2L^2\mathcal R_*''}{\mathcal R_*^3}.
 \label{eq:Vnullsecondgeneral}
\end{equation}
Combining Eqs.~\eqref{eq:generalradialNEC} and
\eqref{eq:Vnullsecondgeneral} yields
\begin{equation}
 V_{\rm null}''\big|_*
 =\frac{8\pi L^2}{\mathcal R_*^2}
 (\rho+p_\ell)_*.
 \label{eq:energyoptical}
\end{equation}
Thus the radial NEC and radial optical stability are locally equivalent:
\begin{align}
 \text{throat}
 &\Longleftrightarrow
 (\rho+p_\ell)_*<0
 \Longleftrightarrow
 \text{unstable light ring},
 \nonumber\\
 \text{anti-throat}
 &\Longleftrightarrow
 (\rho+p_\ell)_*>0
 \Longleftrightarrow
 \text{stable light ring}.
 \nonumber
\end{align}
The qualitative association between wormhole throats, anti-throats, and
extrema of the null potential is present in earlier multi-throat and general
light-ring analyses \cite{Crispim2025,Tsukamoto2024,Xavier2024}.
Equation~\eqref{eq:energyoptical} strengthens the statement within the
ultrastatic spherical class by identifying the local matter combination that
fixes the sign of the optical curvature.

The same relation gives a useful local timescale.  Since $t=E\lambda$ along
a circular null orbit, a small radial perturbation obeys
\begin{equation}
 \frac{d^2\delta\ell}{dt^2}
 +\frac{V_{\rm null}''}{2E^2}\,\delta\ell=0.
 \label{eq:nullperturb}
\end{equation}
Using $E^2=L^2/\mathcal R_*^2$ on the circular orbit,
\begin{equation}
 \frac{V_{\rm null}''}{2E^2}
 =-\frac{\mathcal R_*''}{\mathcal R_*}
 =4\pi(\rho+p_\ell)_*.
 \label{eq:localopticalfrequency}
\end{equation}
Therefore a throat has coordinate-time Lyapunov exponent
\begin{equation}
 \Lambda_{\rm T}^2
 =\frac{\mathcal R_{\rm T}''}{\mathcal R_{\rm T}}
 =-4\pi(\rho+p_\ell)_{\rm T},
 \label{eq:throatLyap}
\end{equation}
whereas an anti-throat has stable small-oscillation frequency
\begin{equation}
 \omega_{\rm A}^2
 =-\frac{\mathcal R_{\rm A}''}{\mathcal R_{\rm A}}
 =4\pi(\rho+p_\ell)_{\rm A}.
 \label{eq:antiFrequency}
\end{equation}
Here $\Lambda_{\rm T}^{-1}$ is the local e-folding time of the unstable
radial displacement, whereas $\omega_{\rm A}^{-1}$ sets the oscillation
timescale of the stable orbit.  This is the familiar geodesic Lyapunov
diagnostic specialized to the proper-distance gauge \cite{Cardoso2009}.

\subsection{Cellwise affine null energy and complete ANEC}
\label{subsec:cellanec}

For a complete affinely parametrized null geodesic $\gamma$, the averaged
null energy condition (ANEC) requires
\begin{equation}
 \mathcal I_\gamma
 \equiv\int_\gamma T_{\mu\nu}k^\mu k^\nu d\lambda\geq0,
 \qquad k^\mu=\frac{dx^\mu}{d\lambda}.
 \label{eq:ANECdefinition}
\end{equation}
Choose a future-directed radial ray running from the left to the right end and
normalize its tangent by $k^t=1$.  In the static orthonormal frame, $k^{\hat\mu}=(1,1,0,0)$.
Thus $d\lambda=d\ell$ and
$T_{\mu\nu}k^\mu k^\nu=\rho+p_\ell$.  Reversing the radial orientation gives
the same averaged sign.  For any finite interval
$[\ell_i,\ell_{i+1}]$ whose endpoints are stationary spheres,
\begin{align}
 \mathcal I_{i,i+1}
 &\equiv
 \int_{\ell_i}^{\ell_{i+1}}
 (\rho+p_\ell)d\ell
 \nonumber\\
 &=-\frac{1}{4\pi}
 \int_{\ell_i}^{\ell_{i+1}}
 \frac{\mathcal R''}{\mathcal R}d\ell.
 \label{eq:cellanecstart}
\end{align}
The quantity $\mathcal I_{i,i+1}$ is a segment integral with dimensions of
inverse length in geometrized units; it is not, by itself, the ANEC integral
on a complete null geodesic.  More generally, integration by parts gives
\begin{equation}
 \int_a^b(\rho+p_\ell)d\ell
 =-\frac{1}{4\pi}\left[\frac{\mathcal R'}{\mathcal R}\right]_a^b
 -\frac{1}{4\pi}\int_a^b
 \left(\frac{\mathcal R'}{\mathcal R}\right)^2d\ell.
 \label{eq:generalANECboundary}
\end{equation}
Using
\begin{equation}
 \frac{d}{d\ell}
 \left(\frac{\mathcal R'}{\mathcal R}\right)
 =\frac{\mathcal R''}{\mathcal R}
 -\left(\frac{\mathcal R'}{\mathcal R}\right)^2
 \label{eq:logderivativegeneral}
\end{equation}
and $\mathcal R'(\ell_i)=\mathcal R'(\ell_{i+1})=0$, one obtains
\begin{equation}
 \mathcal I_{i,i+1}
 =-\frac{1}{4\pi}
 \int_{\ell_i}^{\ell_{i+1}}
 \left(\frac{\mathcal R'}{\mathcal R}\right)^2d\ell<0
 \label{eq:cellanecnegative}
\end{equation}
for every nontrivial cell.  For the complete left-to-right radial null
geodesic, the boundary term also vanishes because
$\mathcal R'/\mathcal R\to0$ at both ends.  We denote the corresponding ANEC
functional by
\begin{equation}
 \begin{aligned}
 \mathcal I_{\rm ANEC}
 &\equiv\int_{-\infty}^{+\infty}
 T_{\mu\nu}k^\mu k^\nu d\lambda
 =\int_{-\infty}^{+\infty}(\rho+p_\ell)d\ell
 \\
 &=-\frac{1}{4\pi}
 \int_{-\infty}^{+\infty}
 \left(\frac{\mathcal R'}{\mathcal R}\right)^2d\ell<0.
 \end{aligned}
 \label{eq:completeanecgeneral}
\end{equation}
provided $\mathcal R'/\mathcal R\to0$ asymptotically.  A constant rescaling of
the affine tangent rescales the numerical value of the integral but not its
sign.

The same argument applies separately to either exterior interval.  Its
boundary term vanishes at the outermost throat and at the corresponding
asymptotic end because $\mathcal R'/\mathcal R\to0$.  Both exterior
contributions are therefore strictly negative as well.

Equation~\eqref{eq:cellanecnegative} provides spatially resolved information
beyond the complete average: every throat--anti-throat or
anti-throat--throat cell has negative integrated radial null energy even
though its anti-throat endpoint satisfies all pointwise energy conditions.
Adding the finite and exterior cells yields the complete two-ended ANEC in
Eq.~\eqref{eq:completeanecgeneral}.

\section{An exactly solvable arbitrary multi-throat hierarchy}
\label{sec:hierarchy}

\subsection{Design criteria and factorized prescription}
\label{subsec:design}

We now solve the inverse problem suggested above: construct a smooth areal
radius with an arbitrary prescribed finite set of stationary spheres while
keeping their classification, asymptotics, and integral analytic.
Reflection symmetry requires $\mathcal R$ to be even and $\mathcal R'$ to be
odd.  Placing symmetric stationary spheres at
$\ell/a=\pm\sqrt{\alpha_j}$ therefore calls for the numerator
$(\ell/a)\prod_j[(\ell/a)^2-\alpha_j]$, whose degree is $2N+1$, where $ 0<\alpha_1<\alpha_2<\cdots<\alpha_N $ are dimensionless parameters.
In keeping with the geometrical interpretation, radial
dependence is displayed explicitly through the dimensionless quantity
$\ell/a$ rather than through an auxiliary radial coordinate.

Two-ended asymptotic flatness in proper distance requires
$\mathcal R'\to\pm1$.  A positive, zero-free denominator with the same
large-$|\ell|$ degree is consequently needed.  The minimal smooth choice
$[1+(\ell/a)^2]^{N+1/2}$ introduces no additional real zeros or poles, fixes
the asymptotic normalization without a further parameter, and makes the
primitive a finite linear combination of half-integer powers of
$1+(\ell/a)^2$.  The resulting prescription is not unique, but it is the
simplest member of a broad design class satisfying all four requirements at
once: prescribed roots, global regularity, unit asymptotic slopes, and
elementary integrability.

Let $N\in\mathbb N_0\equiv\{0,1,2,\ldots\}$ label the order of the
hierarchy.  For $N\geq1$, with $ 0<\alpha_1<\alpha_2<\cdots<\alpha_N $ dimensionless parameters and choose
\begin{equation}
 \mathcal R_N'(\ell)
 =\frac{(\ell/a)
 \displaystyle\prod_{j=1}^{N}
 \left[(\ell/a)^2-\alpha_j\right]}
 {\left[1+(\ell/a)^2\right]^{N+1/2}}.
 \label{eq:hierarchyderivative}
\end{equation}
The length $a>0$ fixes the overall scale.  Equation~\eqref{eq:hierarchyderivative}
is odd in $\ell$, so the integrated areal radius is even up to an additive
constant.  The ordered roots control the proper-distance architecture,
whereas that constant controls the absolute areal size without moving any
stationary sphere.  This clean separation between location data and size
data is a useful feature of the construction.

Define $ \beta_j\equiv1+\alpha_j$,
and let $e_k(\beta_1,\ldots,\beta_N)$ denote the elementary symmetric
polynomial of degree $k$, with $e_0=1$.  Direct integration gives
\begin{eqnarray}
	\frac{\mathcal R_N(\ell)}{a}
	&=& C_N+\sqrt{1+(\ell/a)^2}
		\nonumber \\
	&&+\sum_{k=1}^{N}
	\frac{(-1)^{k+1}e_k(\beta_1,\ldots,\beta_N)}{2k-1}\times
		\nonumber \\
	&&\qquad \times \left[1+(\ell/a)^2\right]^{-(k-1/2)}.
	\label{eq:hierarchyclosed}
\end{eqnarray}
%
To see why the result terminates, expand the numerator as
\begin{equation}
 \prod_{j=1}^{N}
 \left[(\ell/a)^2-\alpha_j\right]
 =\prod_{j=1}^{N}
 \left(1+(\ell/a)^2-\beta_j\right).
 \label{eq:appendixproduct}
\end{equation}
The elementary-symmetric-polynomial expansion reduces the integrand to a
finite sequence of powers of $1+(\ell/a)^2$.  Term-by-term integration then
produces
$[1+(\ell/a)^2]^{1/2},[1+(\ell/a)^2]^{-1/2},\ldots,$ exactly as displayed
in Eq.~\eqref{eq:hierarchyclosed}.  Thus the closed form is a direct
consequence of the same denominator that enforces the correct asymptotic
slope; analytical solvability and asymptotic flatness are built into a
single choice.

Here $C_N$ is a dimensionless integration constant.  To state its role
precisely, write Eq.~\eqref{eq:hierarchyclosed} as
\begin{equation}
 \frac{\mathcal R_N}{a}=C_N+F_N(x),
 \qquad x\equiv\frac{\ell}{a},
 \label{eq:FNdefinition}
\end{equation}
where $F_N$ denotes the complete nonconstant part of the closed expression.
Then global positivity is guaranteed by the strict bound
\begin{equation}
 C_N>-\min_{x\in\mathbb R}F_N(x).
 \label{eq:CNpositivity}
\end{equation}
Because $F_N\to+\infty$ as $|x|\to\infty$, its minimum occurs at one of the
finite stationary points $x=0,\pm\sqrt{\alpha_j}$.  The constant $C_N$ does
not move or reclassify these spheres, because it leaves $\mathcal R_N'$ and
$\mathcal R_N''$ unchanged.  It is nevertheless physically relevant: by
shifting the areal radii, it changes areas, curvature ratios, the effective
stress tensor, circular-orbit impact parameters, and optical timescales.

The asymptotic expansion is
\begin{equation}
 \mathcal R_N(\ell)
 =|\ell|+aC_N
 +\frac{a^2\mathcal A_N}{|\ell|}
 +O(|\ell|^{-3}),
 \label{eq:hierarchyasymR}
\end{equation}
where
\begin{equation}
 \mathcal A_N
 \equiv N+\frac12+\sum_{j=1}^{N}\alpha_j>0.
 \label{eq:ANdef}
\end{equation}
Thus $\mathcal R_N'\to\pm1$ and every member is asymptotically flat on both
sides.  The Misner--Sharp mass \cite{MisnerSharp1964},
\begin{equation}
 m_{\rm MS}=\frac{\mathcal R_N}{2}\left(1-\mathcal R_N'^2\right),
 \label{eq:hierarchyMS}
\end{equation}
takes the asymptotic form
\begin{equation}
 m_{\rm MS}=\frac{\mathcal A_Na^2}{|\ell|}
 +O(|\ell|^{-2}),
 \label{eq:hierarchyMSasym}
\end{equation}
and therefore tends to zero.  Equivalently, $1-\mathcal R_N'^2$ has no
$1/\mathcal R_N$ term, so the Arnowitt--Deser--Misner (ADM) mass vanishes.
This conclusion follows from the particular radial falloff in
Eq.~\eqref{eq:hierarchyasymR}, not from a unit lapse by itself.  The leading source terms are
\begin{align}
 \rho&=-\frac{\mathcal A_Na^2}{4\pi\ell^4}
 +O(|\ell|^{-5}),
 \\
 p_\ell&=-\frac{\mathcal A_Na^2}{4\pi\ell^4}
 +O(|\ell|^{-5}),
 \\
 p_t&=+\frac{\mathcal A_Na^2}{4\pi\ell^4}
 +O(|\ell|^{-5}).
 \label{eq:hierarchysourceasym}
\end{align}
Hence the hierarchy retains the fast $|\ell|^{-4}$ source falloff of the
$N=1$ example.
The leading tail is anisotropic: $\rho$ and $p_\ell$ are negative and equal,
whereas $p_t$ has the opposite sign.  Nevertheless, the falloff is fast
enough that no $1/\mathcal R_N$ mass term survives at infinity.  The
vanishing ADM mass is therefore compatible with nonzero local curvature and
stress energy in the interior.

\subsection{Exact stationary-sphere classification}
\label{subsec:hierarchyclassification}

Equation~\eqref{eq:hierarchyderivative} makes the stationary locations
manifest:
\begin{equation}
 \ell=0,
 \qquad
 \frac{\ell}{a}=\pm\sqrt{\alpha_k},
 \qquad k=1,\ldots,N.
 \label{eq:hierarchystationary}
\end{equation}
The second derivative at the center is
\begin{equation}
 \mathcal R_N''(0)
 =\frac{(-1)^N}{a}
 \prod_{j=1}^{N}\alpha_j.
 \label{eq:hierarchysecondcenter}
\end{equation}
At the nonzero stationary spheres,
\begin{equation}
 \mathcal R_N''\!\left(\pm a\sqrt{\alpha_k}\right)
 =\frac{2\alpha_k
 \displaystyle\prod_{j\neq k}(\alpha_k-\alpha_j)}
 {a(1+\alpha_k)^{N+1/2}}.
 \label{eq:hierarchysecondroots}
\end{equation}
Because the $\alpha_j$ are ordered,
\begin{equation}
 \operatorname{sgn}
 \mathcal R_N''\!\left(\pm a\sqrt{\alpha_k}\right)
 =(-1)^{N-k}.
 \label{eq:hierarchysignroots}
\end{equation}
The extrema therefore alternate automatically.

For odd $N$, the center is an anti-throat; the roots with odd $k$ are
throats and those with even $k$ are anti-throats.  For even $N$, the center
is a throat; the roots with odd $k$ are anti-throats and those with even $k$
are throats.  In either case,
\begin{equation}
 N_{\rm T}=N+1,
 \qquad
 N_{\rm A}=N,
 \label{eq:hierarchycount}
\end{equation}
in exact agreement with the general counting relation
\eqref{eq:generalcount}.  The parameters $a\sqrt{\alpha_k}$ directly set the
proper-distance locations of the stationary spheres.  In particular, the
proper separation between two neighboring positive-$\ell$ extrema is
\begin{equation}
 \Delta\ell_{k,k+1}
 =a\left(\sqrt{\alpha_{k+1}}-\sqrt{\alpha_k}\right),
 \label{eq:neighborseparation}
\end{equation}
while the center-to-first-extremum distance is $a\sqrt{\alpha_1}$.

\begin{figure}[t]
 \centering
 \includegraphics[width=\columnwidth]{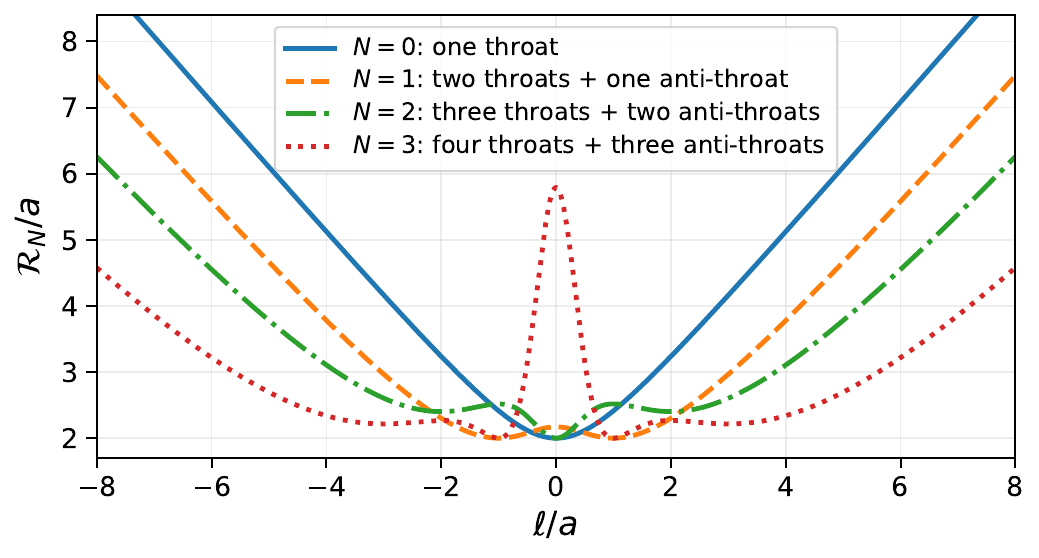}
 \caption{Representative members of the exact hierarchy.  The $N=0$ member
 has one throat.  The $N=1$ member has two throats and one anti-throat.  The
 $N=2$ member has three throats and two anti-throats.  The $N=3$ member has
 four throats and three anti-throats.  For $N\ge1$ the illustrative choices
 are $\{\alpha_j\}=\{1\}$, $\{1,4\}$, and $\{1,4,9\}$, respectively.  The
 constants $C_N$ are shifted only to set the smallest radius equal to $2a$;
 such shifts do not affect any stationary location or its classification.}
 \label{fig:hierarchy}
\end{figure}

Figure~\ref{fig:hierarchy} shows the first four multiplicities.  For visual
comparison the integration constants $C_N$ have been chosen so
that the smallest areal radius of each plotted profile is $2a$.  The figure
is not a sequence generated by varying a single parameter; it is a visual
summary of the exact hierarchy $N=0,1,2,3$.

\subsection{Low-order members and the Ellis limit}
\label{subsec:lowordermembers}

For $N=0$, Eq.~\eqref{eq:hierarchyclosed} gives
\begin{equation}
 \frac{\mathcal R_0}{a}
 =C_0+\sqrt{1+(\ell/a)^2}.
 \label{eq:N0}
\end{equation}
With $C_0=0$ this is the standard Ellis--Bronnikov areal radius
\cite{Ellis1973,Bronnikov1973}.

For $N=1$, writing $\alpha_1\equiv\alpha$ and choosing $C_1=0$ gives
\begin{equation}
 \frac{\mathcal R_1(\ell)}{a}
 =\frac{(\ell/a)^2+\alpha+2}
 {\sqrt{1+(\ell/a)^2}}.
 \label{eq:N1profile}
\end{equation}
This is the double-throat solution analyzed in detail in
Sec.~\ref{sec:N1}.  Its analytic continuation away from the hierarchy range
$\alpha>0$ is useful: for $-2<\alpha<0$ it has one throat, at $\alpha=0$
the throat is quartically degenerate, and for $\alpha>0$ it contains two
throats separated by a central anti-throat.  At $\alpha=-1$ it reduces
exactly to Eq.~\eqref{eq:N0} with $C_0=0$.

For $N=2$,
\begin{equation}
 \begin{aligned}
 \frac{\mathcal R_2}{a}
 =&\,C_2+\sqrt{1+(\ell/a)^2}
 +\frac{\beta_1+\beta_2}{\sqrt{1+(\ell/a)^2}}
 \\
 &-\frac{\beta_1\beta_2}
 {3\left[1+(\ell/a)^2\right]^{3/2}}.
 \end{aligned}
 \label{eq:N2explicit}
\end{equation}
Its stationary pattern is
\begin{equation}
 \begin{array}{ccl}
 \ell=0,\ \ell/a=\pm\sqrt{\alpha_2} &:& \text{throats},\\
 \ell/a=\pm\sqrt{\alpha_1} &:& \text{anti-throats}.
 \end{array}
 \label{eq:N2pattern}
\end{equation}
This is the simplest member with three throats, two anti-throats, and two
finite optical cavities.

For general $N=3$, define
\begin{align}
 e_1&=\beta_1+\beta_2+\beta_3,
 \nonumber\\
 e_2&=\beta_1\beta_2+\beta_1\beta_3+\beta_2\beta_3,
 \nonumber\\
 e_3&=\beta_1\beta_2\beta_3.
 \nonumber
\end{align}
Then
\begin{equation}
 \begin{aligned}
 \frac{\mathcal R_3}{a}
 =&\,C_3+\sqrt{1+(\ell/a)^2}
 +\frac{e_1}{\sqrt{1+(\ell/a)^2}}
 \\
 &-\frac{e_2}{3\left[1+(\ell/a)^2\right]^{3/2}}
 +\frac{e_3}{5\left[1+(\ell/a)^2\right]^{5/2}}.
 \end{aligned}
 \label{eq:N3generalappendix}
\end{equation}
The center and the $k=2$ pair are anti-throats, whereas the $k=1$ and $k=3$
pairs are throats.  The following concrete choice makes the resulting
three-cavity chain particularly transparent.

For $N=3$ and $ \alpha_1=1$, $ \alpha_2=4$, $\alpha_3=9$,
Eq.~\eqref{eq:hierarchyclosed} becomes
\begin{equation}
 \begin{aligned}
 \frac{\mathcal R_3}{a}
 =&\,C_3+\sqrt{1+(\ell/a)^2}
 +\frac{17}{\sqrt{1+(\ell/a)^2}}
 \\
 &-\frac{80}{3\left[1+(\ell/a)^2\right]^{3/2}}
 +\frac{20}{\left[1+(\ell/a)^2\right]^{5/2}}.
 \end{aligned}
 \label{eq:N3explicit}
\end{equation}
Its stationary spheres lie at
\begin{equation}
 \frac{\ell}{a}=0,\ \pm1,\ \pm2,\ \pm3,
 \label{eq:N3locations}
\end{equation}
with the exact pattern
\begin{equation}
 \begin{array}{ccl}
 \ell/a=0,\ \pm2 &:& \text{anti-throats},\\
 \ell/a=\pm1,\ \pm3 &:& \text{throats}.
 \end{array}
 \label{eq:N3classification}
\end{equation}
Thus the spacetime has four necks and three finite islands.  The central
island is bounded by the throats at $\ell/a=\pm1$, while two additional
islands lie between the pairs $(-3,-1)$ and $(1,3)$.

\section{The \texorpdfstring{$N=1$}{N=1} double-throat member in detail}
\label{sec:N1}

We now specialize to Eq.~\eqref{eq:N1profile}.  This worked example is useful
for three reasons.  First, all relevant quantities remain elementary.
Second, it provides a continuous one-parameter interpolation between the
Ellis--Bronnikov geometry and a double-throat configuration.  Third, it
allows direct comparison with the recent two-throat literature while the
general hierarchy clarifies which conclusions survive beyond two throats.

The first two derivatives are
\begin{align}
 \mathcal R_1'
 &=\frac{(\ell/a)\left[(\ell/a)^2-\alpha\right]}
 {\left[1+(\ell/a)^2\right]^{3/2}},
 \label{eq:N1Rprime}
 \\
 \mathcal R_1''
 &=\frac{(3+2\alpha)(\ell/a)^2-\alpha}
 {a\left[1+(\ell/a)^2\right]^{5/2}}.
 \label{eq:N1Rsecond}
\end{align}
For $\alpha>0$, the throats are at
\begin{equation}
 \frac{\ell_{\rm T}^{\pm}}{a}=\pm\sqrt{\alpha},
 \label{eq:N1throatlocations}
\end{equation}
and the central anti-throat is at $\ell=0$.  Their areal radii are
\begin{align}
 \mathcal R_{\rm T}&=2a\sqrt{1+\alpha},
 \label{eq:N1RT}
 \\
 \mathcal R_{\rm A}&=a(2+\alpha).
 \label{eq:N1RA}
\end{align}
The bulge height and proper throat separation are
\begin{align}
 \mathcal R_{\rm A}-\mathcal R_{\rm T}
 &=a\left(\sqrt{1+\alpha}-1\right)^2,
 \label{eq:N1bulgeheight}
 \\
 \Delta\ell_{\rm T}&=2a\sqrt{\alpha}.
 \label{eq:N1throatseparation}
\end{align}
Writing $A_{\rm T}=4\pi\mathcal R_{\rm T}^2$ and
$A_{\rm A}=4\pi\mathcal R_{\rm A}^2$ for the throat and anti-throat areas,
respectively, the scale-free area enhancement is
\begin{equation}
 \frac{A_{\rm A}}{A_{\rm T}}
 =1+\frac{\alpha^2}{4(1+\alpha)}.
 \label{eq:N1arearatio}
\end{equation}

\subsection{Single-to-double-throat bifurcation}
\label{subsec:N1bifurcation}

\begin{figure}[t]
 \centering
 \includegraphics[width=\columnwidth]{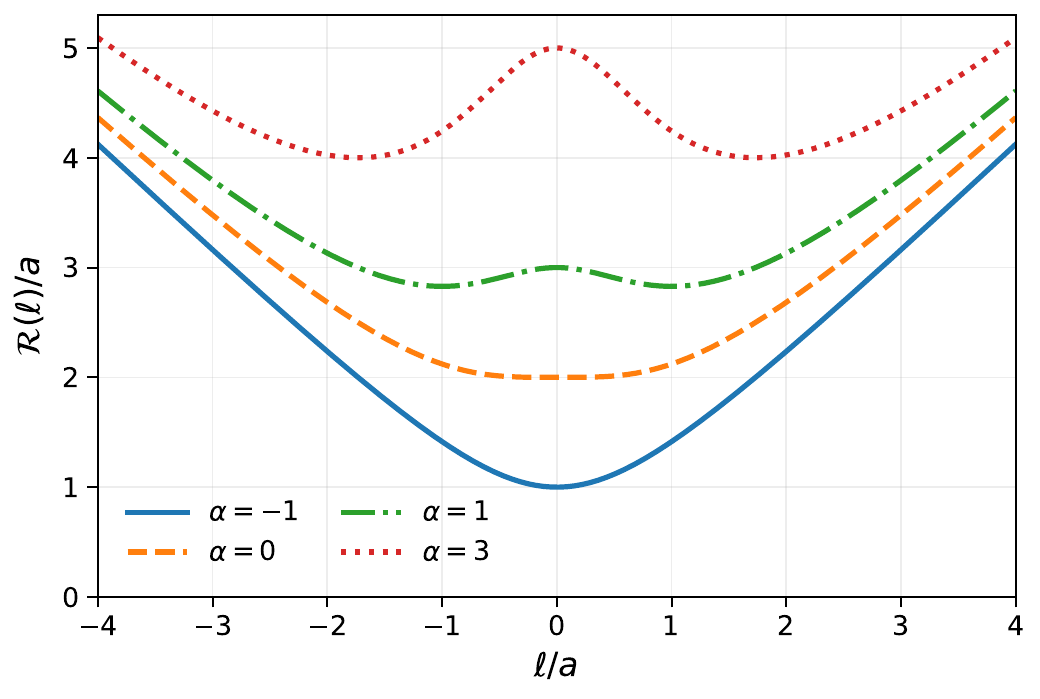}
 \caption{Dimensionless areal radius $\mathcal R_1/a$ versus $\ell/a$ for
 representative values of $\alpha$.  The Ellis--Bronnikov member
 $\alpha=-1$ has one throat.  At $\alpha=0$ the center is quartically
 degenerate.  For $\alpha>0$ the central surface becomes an anti-throat and
 two throats appear at $\ell/a=\pm\sqrt{\alpha}$.}
 \label{fig:N1profiles}
\end{figure}

For $-2<\alpha<0$, the only stationary sphere is the center and
$\mathcal R_1''(0)>0$, so the geometry has a single throat.  At
$\alpha=-1$,
\begin{equation}
 \mathcal R_1(\ell)=\sqrt{\ell^2+a^2},
 \label{eq:N1Ellis}
\end{equation}
recovering the massless Ellis--Bronnikov wormhole.  Expanding near the
center gives
\begin{align}
 \frac{\mathcal R_1}{a}
 =&\,2+\alpha
 -\frac{\alpha}{2}\left(\frac{\ell}{a}\right)^2
 \nonumber\\
 &+\left(\frac14+\frac{3\alpha}{8}\right)
 \left(\frac{\ell}{a}\right)^4
 +O\!\left[(\ell/a)^6\right].
 \label{eq:N1localexpansion}
\end{align}
At the critical value $\alpha=0$,
\begin{equation}
 \left.\frac{\mathcal R_1}{a}\right|_{\alpha=0}
 =2+\frac14\left(\frac{\ell}{a}\right)^4
 +O\!\left[(\ell/a)^6\right],
 \label{eq:N1quartic}
\end{equation}
so the central throat is quartically degenerate.  For $\alpha>0$ the
quadratic term changes sign and the center becomes an anti-throat while two
symmetric minima emerge.  Figure~\ref{fig:N1profiles} displays this
bifurcation.

The transition is also an optical bifurcation.  At $\alpha=0$ the central
null potential has a quartically flat maximum rather than the ordinary
quadratic maximum; for $\alpha>0$ it becomes a stable minimum flanked by two
unstable maxima.  Marginal photon spheres require a strong-deflection
expansion different from the standard nondegenerate logarithmic case
\cite{Igata2026}.  The exactly solvable critical member therefore provides a
simple test background for studying how lensing and scattering observables
change across light-ring creation.

\subsection{Proper size and visualization of the central island}
\label{subsec:N1island}

The proper three-volume between the two throats is
\begin{equation}
 V_{\rm island}
 =4\pi\int_{-a\sqrt{\alpha}}^{+a\sqrt{\alpha}}
 \mathcal R_1^2(\ell)d\ell,
 \label{eq:N1islandvolumeDef}
\end{equation}
which evaluates to
\begin{equation}
 \begin{aligned}
 V_{\rm island}=8\pi a^3\bigg[&
 \sqrt{\alpha}\left(3+\frac{7\alpha}{3}\right)
 \\
 &+(1+\alpha)^2\arctan\sqrt{\alpha}\bigg].
 \end{aligned}
 \label{eq:N1islandvolume}
\end{equation}
The limiting behavior is
\begin{align}
 V_{\rm island}
 &=32\pi a^3\sqrt{\alpha}+O(\alpha^{3/2}),
 \qquad \alpha\to0^+,
 \label{eq:N1volsmall}
 \\
 V_{\rm island}
 &=4\pi^2a^3\alpha^2+O(\alpha^{3/2}),
 \qquad \alpha\to+\infty.
 \label{eq:N1vollarge}
\end{align}
The large-$\alpha$ scaling is worth noting.  The throat separation grows as
$a\sqrt{\alpha}$, while the central maximum radius grows as $a\alpha$; the
region retaining the largest radius has proper thickness of order $a$, so
the leading volume scales as $a^3\alpha^2$ rather than $a^3\alpha^{5/2}$.

\begin{figure}[t]
 \centering
 \includegraphics[width=\columnwidth]{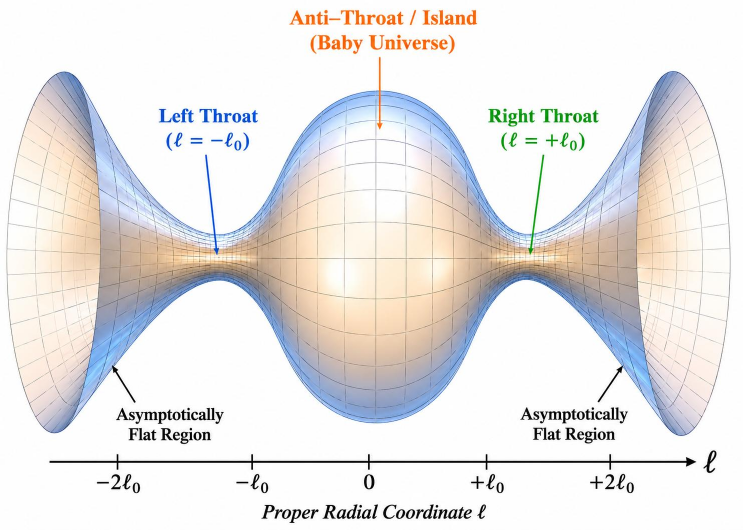}
 \caption{Schematic three-dimensional embedding-style visualization of the $N=1$
 double-throat geometry.  Here $\ell_0\equiv a\sqrt{\alpha}$.  Moving from
 either asymptotically flat end, the spatial section narrows to a throat,
 opens into an enlarged central chamber, reaches its largest symmetry sphere
 at the anti-throat, narrows through the second throat, and opens into the
 opposite asymptotic region.  The central bulge is therefore
 ``baby-universe-like'' only in a morphological sense; the label in the
 schematic does not denote a third asymptotic region, a disconnected daughter
 universe, or a dynamical creation process.}
 \label{fig:N1embedding3d}
\end{figure}

Figure~\ref{fig:N1embedding3d} gives a schematic embedding-style visualization
of the throat--island--throat morphology.  It is intended only as a
geometrical aid: it is not generated from a specified value of $\alpha$ and
is not a substitute for the exact Euclidean embedding analysis in
Sec.~\ref{subsec:N1embedding}.

\subsection{Branch-resolved Morris--Thorne representation}
\label{subsec:N1shape}

The areal radius is not a global coordinate on the double-throat branch.
Nevertheless, the Morris--Thorne shape function can be recovered on each
monotonic segment.  Define
\begin{equation}
 Y\equiv\sqrt{\ell^2+a^2},
 \qquad
 q^2\equiv a^2(1+\alpha).
 \label{eq:N1Yq}
\end{equation}
Then
\begin{equation}
 \mathcal R_1=Y+\frac{q^2}{Y}.
 \label{eq:N1RY}
\end{equation}
Solving for $Y$ gives
\begin{equation}
 Y_\pm(\mathcal R_1)
 =\frac{\mathcal R_1
 \pm\sqrt{\mathcal R_1^2-4q^2}}{2}.
 \label{eq:N1Ybranches}
\end{equation}
The $Y_+$ branch describes either exterior beyond a throat, while $Y_-$
describes the central segment between a throat and the anti-throat.  Since
\begin{equation}
 \left(\frac{d\mathcal R_1}{d\ell}\right)^2
 =\frac{(Y^2-a^2)(Y^2-q^2)^2}{Y^6},
 \label{eq:N1RprimeY}
\end{equation}
the branchwise Morris--Thorne form
\begin{equation}
 ds^2=-dt^2+
 \frac{d\mathcal R_1^2}{1-b_{\rm sh}(\mathcal R_1)/\mathcal R_1}
 +\mathcal R_1^2d\Omega^2
 \label{eq:N1MTform}
\end{equation}
has
\begin{equation}
 b_{{\rm sh},\pm}(\mathcal R_1)
 =\mathcal R_1\left[
 1-\frac{(Y_\pm^2-a^2)(Y_\pm^2-q^2)^2}{Y_\pm^6}
 \right].
 \label{eq:N1shapeexplicit}
\end{equation}
Here $b_{\rm sh}$ denotes the Morris--Thorne shape function; the subscript
prevents confusion with an optical impact parameter introduced below.  At any
stationary sphere $b_{\rm sh}=\mathcal R_1$, so this equation alone does not
distinguish a throat from an anti-throat.  Differentiating
$1-b_{\rm sh}/\mathcal R_1=\mathcal R_1'^2$ and taking the regular branch
limit gives
\begin{equation}
 \left.\frac{db_{\rm sh}}{d\mathcal R_1}\right|_*
 =1-2\mathcal R_*\mathcal R_*''.
 \label{eq:N1bprimegeneral}
\end{equation}
Hence
\begin{align}
 \left.\frac{db_{\rm sh}}{d\mathcal R_1}\right|_{\rm T}
 &=\frac{1-7\alpha}{1+\alpha}<1,
 \label{eq:N1bprimeT}
 \\
 \left.\frac{db_{\rm sh}}{d\mathcal R_1}\right|_{\rm A}
 &=1+2\alpha(\alpha+2)>1.
 \label{eq:N1bprimeA}
\end{align}
The usual flare-out inequality holds at the two minima and reverses at the
central maximum.

\subsection{Exact source and pointwise energy conditions}
\label{subsec:N1source}

Substitution of Eqs.~\eqref{eq:N1profile}, \eqref{eq:N1Rprime}, and
\eqref{eq:N1Rsecond} into the
universal Einstein equations gives
\begin{align}
 8\pi a^2\rho
 =&\,
 \frac{1+4\alpha+2\alpha^2
 -(9+12\alpha+5\alpha^2)(\ell/a)^2}
 {\left[1+(\ell/a)^2\right]^2
 \left[\alpha+(\ell/a)^2+2\right]^2}
 \nonumber\\
 &-\frac{(3+2\alpha)(\ell/a)^4}
 {\left[1+(\ell/a)^2\right]^2
 \left[\alpha+(\ell/a)^2+2\right]^2},
 \label{eq:N1rho}
 \\
 8\pi a^2p_\ell
 =&\,
 \frac{-1+(\alpha^2-3)(\ell/a)^2
 -(3+2\alpha)(\ell/a)^4}
 {\left[1+(\ell/a)^2\right]^2
 \left[\alpha+(\ell/a)^2+2\right]^2},
 \label{eq:N1pl}
 \\
 8\pi a^2p_t
 =&\,
 \frac{(3+2\alpha)(\ell/a)^2-\alpha}
 {\left[1+(\ell/a)^2\right]^2
 \left[\alpha+(\ell/a)^2+2\right]}.
 \label{eq:N1pt}
\end{align}
At $\alpha=-1$ these expressions reduce to
\begin{equation}
 \rho=p_\ell
 =-\frac{a^2}{8\pi(\ell^2+a^2)^2},
 \qquad
 p_t=+\frac{a^2}{8\pi(\ell^2+a^2)^2},
 \label{eq:N1Ellissource}
\end{equation}
which is the familiar Ellis--Bronnikov ghost-scalar stress pattern. Appendix~\ref{app:source} reconstructs the full $N=1$ stress tensor, and more generally that of the hierarchy, using a scalar with a field-dependent kinetic coefficient and a scalar-dependent electromagnetic coupling. The energy conditions below refer to the total source, independently of this particular decomposition.

\begin{figure}[t]
 \centering
 \includegraphics[width=\columnwidth]{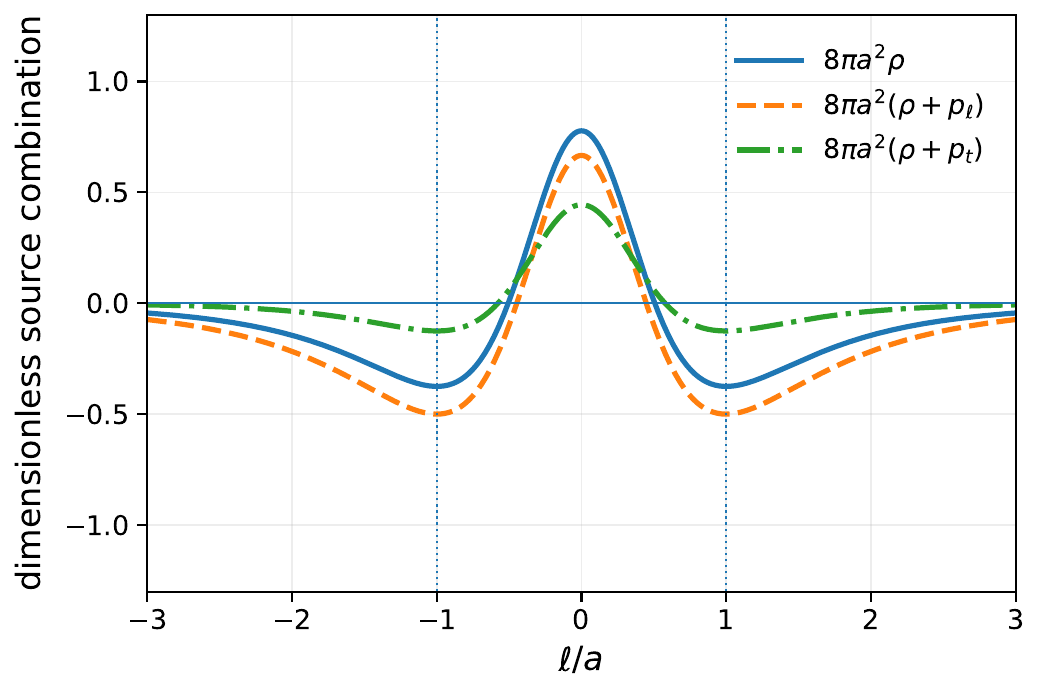}
 \caption{Representative source and NEC combinations for the $N=1$ member
 with $\alpha=1$.  The vertical dotted lines mark the throats at
 $\ell/a=\pm1$.  Both NEC combinations and the energy density are positive
 at the central anti-throat, whereas the radial NEC becomes negative between the central anti-throat and each throat and is necessarily violated at both necks.}
 \label{fig:N1energy}
\end{figure}

The independent NEC combinations simplify to
\begin{equation}
 8\pi a^2(\rho+p_\ell)
 =\frac{2\left[\alpha-(3+2\alpha)(\ell/a)^2\right]}
 {\left[1+(\ell/a)^2\right]^2
 \left[\alpha+(\ell/a)^2+2\right]}
 \label{eq:N1NECr}
\end{equation}
and
\begin{equation}
 8\pi a^2(\rho+p_t)
 =\frac{(1+\alpha)^2\left[1-3(\ell/a)^2\right]}
 {\left[1+(\ell/a)^2\right]^2
 \left[\alpha+(\ell/a)^2+2\right]^2}.
 \label{eq:N1NECt}
\end{equation}
On the double-throat branch $\alpha>0$, the NEC holds in all null directions in the closed central interval
\begin{equation}
 \left|\frac{\ell}{a}\right|\leq
 \min\left\{
 \sqrt{\frac{\alpha}{3+2\alpha}},\frac{1}{\sqrt3}
 \right\}.
 \label{eq:N1NECcore}
\end{equation}
Both inequalities are strict in the interior, and the controlling boundary changes at $\alpha=3$. Figure~\ref{fig:N1energy} shows the source for $\alpha=1$.

At the anti-throat,
\begin{align}
 8\pi a^2\rho_{\rm A}
 &=\frac{1+4\alpha+2\alpha^2}{(\alpha+2)^2},
 \\
 8\pi a^2p_{\ell\rm A}
 &=-\frac{1}{(\alpha+2)^2},
 \\
 8\pi a^2p_{t\rm A}
 &=-\frac{\alpha}{\alpha+2},
 \label{eq:N1antiValues}
\end{align}
and all standard pointwise energy conditions hold, in accordance with the
universal theorem of Sec.~\ref{subsec:energytheorem}.  At either throat,
\begin{align}
 8\pi a^2\rho_{\rm T}
 &=\frac{1-7\alpha}{4(1+\alpha)^2},
 \\
 8\pi a^2p_{\ell\rm T}
 &=-\frac{1}{4(1+\alpha)},
 \\
 8\pi a^2(\rho+p_\ell)_{\rm T}
 &=-\frac{2\alpha}{(1+\alpha)^2}<0.
 \label{eq:N1throatValues}
\end{align}
The throat density is nevertheless positive for $0<\alpha<1/7$; in that
range
\begin{equation}
 w_{\ell\rm T}
 =\frac{p_{\ell\rm T}}{\rho_{\rm T}}
 =-\frac{1+\alpha}{1-7\alpha}<-1.
 \label{eq:N1throatw}
\end{equation}
Thus NEC violation, rather than negative density by itself, is the invariant
flare-out requirement.
This is physically useful because it separates two notions that are often
conflated: a positive local energy density can coexist with the radial
tension needed to violate the NEC.  In the narrow range
$0<\alpha<1/7$, the double-throat geometry realizes this distinction
explicitly at both necks.

\subsection{Exact averaged exoticity}
\label{subsec:N1averaged}

The general identity \eqref{eq:completeanecgeneral} can be integrated in
closed form for the $N=1$ profile.  With the normalization
$k^t=1$, $\mathcal I_{\rm ANEC}$ denotes the complete affine line integral
defined in Eq.~\eqref{eq:completeanecgeneral}; it is not a spatial volume
integral.  For the double-throat branch $\alpha>0$,
\begin{equation}
 \mathcal I_{\rm ANEC}
 =-\frac{1}{8a(1+\alpha)}
 \left[
 \alpha+9-\frac{4(\alpha+3)}{\sqrt{\alpha+2}}
 \right]<0.
 \label{eq:N1ANECexact}
\end{equation}
The central NEC-respecting island therefore cannot compensate for the net
negative affine radial null energy. The same expression extends continuously to the single-throat and critical branches, with a removable singularity at $\alpha=-1$. Useful checks are
\begin{align}
 \lim_{\alpha\to-1}\mathcal I_{\rm ANEC}
 &=-\frac{1}{8a},
 \qquad\text{(Ellis--Bronnikov limit)},
 \nonumber\\
 \left.\mathcal I_{\rm ANEC}\right|_{\alpha=0}
 &=-\frac{9-6\sqrt2}{8a},
 \nonumber\\
 \mathcal I_{\rm ANEC}
 &\longrightarrow-\frac{1}{8a},
 \qquad \alpha\to+\infty.
 \label{eq:N1ANEClimits}
\end{align}
The first limit is removable: both the numerator and denominator of
Eq.~\eqref{eq:N1ANECexact} vanish at $\alpha=-1$.  The overall $a^{-1}$
scaling reflects the inverse-length dimension of the affine integral.

A distinct diagnostic, motivated by volume-integral quantifiers of the amount
and distribution of energy-condition violation
\cite{VisserKarDadhich2003,KarDadhichVisser2004}, is obtained here using the
proper spatial measure.  On a constant-time slice the proper volume element
is $dV_{\rm prop}=4\pi\mathcal R_1^2d\ell$, and we define
\begin{equation}
 \mathcal Q_V
 \equiv\int_{-\infty}^{+\infty}(\rho+p_\ell)dV_{\rm prop}
 =4\pi\int_{-\infty}^{+\infty}
 (\rho+p_\ell)\mathcal R_1^2d\ell,
 \label{eq:N1QVdef}
\end{equation}
which evaluates to
\begin{equation}
 \mathcal Q_V
 =\frac{\pi a}{8}\left(\alpha^2-6\alpha-15\right).
 \label{eq:N1QVexact}
\end{equation}
Hence $\mathcal Q_V$ changes sign at $ \alpha=3+2\sqrt6$.
This does not conflict with Eq.~\eqref{eq:N1ANECexact}: the two quantities
use different geometrical weightings and even have different dimensions:
$\mathcal I_{\rm ANEC}$ scales as inverse length, while $\mathcal Q_V$ scales
as length.  The affine null average is tied to null focusing and remains
negative, whereas a sufficiently large positive radial-null-energy contribution from the central region can dominate a spatial proper-volume measure.  This measure-dependence is
consistent with quantum-inequality and exoticity analyses
\cite{FordRoman1996,FewsterRoman2005}.

\subsection{Exact Euclidean embedding criterion}
\label{subsec:N1embedding}

On a $t=\mathrm{const}$, $\theta=\pi/2$ slice,
\begin{equation}
 d\sigma^2=d\ell^2+\mathcal R_1^2d\phi^2.
 \label{eq:N1equatorialslice}
\end{equation}
Embedding the slice as a surface of revolution in Euclidean three-space
requires
\begin{equation}
 \left(\frac{dz}{d\ell}\right)^2
 =1-\mathcal R_1'^2.
 \label{eq:N1embedcondition}
\end{equation}
For the present profile,
\begin{equation}
 1-\mathcal R_1'^2
 =\frac{1+(3-\alpha^2)(\ell/a)^2
 +(3+2\alpha)(\ell/a)^4}
 {\left[1+(\ell/a)^2\right]^3}.
 \label{eq:N1embedradicand}
\end{equation}
The discriminant of the numerator, regarded as a quadratic polynomial in
$(\ell/a)^2$, factorizes as
\begin{equation}
 \Delta=(\alpha-3)(\alpha+1)^3.
 \label{eq:N1embeddisc}
\end{equation}
Thus the strict double-throat slice admits a global isometric realization as a $C^1$ surface of revolution in ordinary Euclidean three-space precisely for $ 0<\alpha\le3$.
For $0<\alpha<3$ the monotonic embedding is smooth. At $\alpha=3$, the absolute-value prescription below gives a $C^1$ meridional curve whose second derivative is discontinuous at $\ell/a=\pm1/\sqrt3$; the endpoint is therefore included with this regularity qualification. The endpoint $\alpha=0$ is also globally embeddable, but it is the
quartically degenerate critical geometry rather than a strict double-throat
member.
For $\alpha>3$ the spacetime remains smooth and asymptotically flat; only the
representation of the entire equatorial slice as a Euclidean surface of
revolution fails over regions where $|\mathcal R_1'|>1$.

\begin{figure}[t]
 \centering
 \includegraphics[width=\columnwidth]{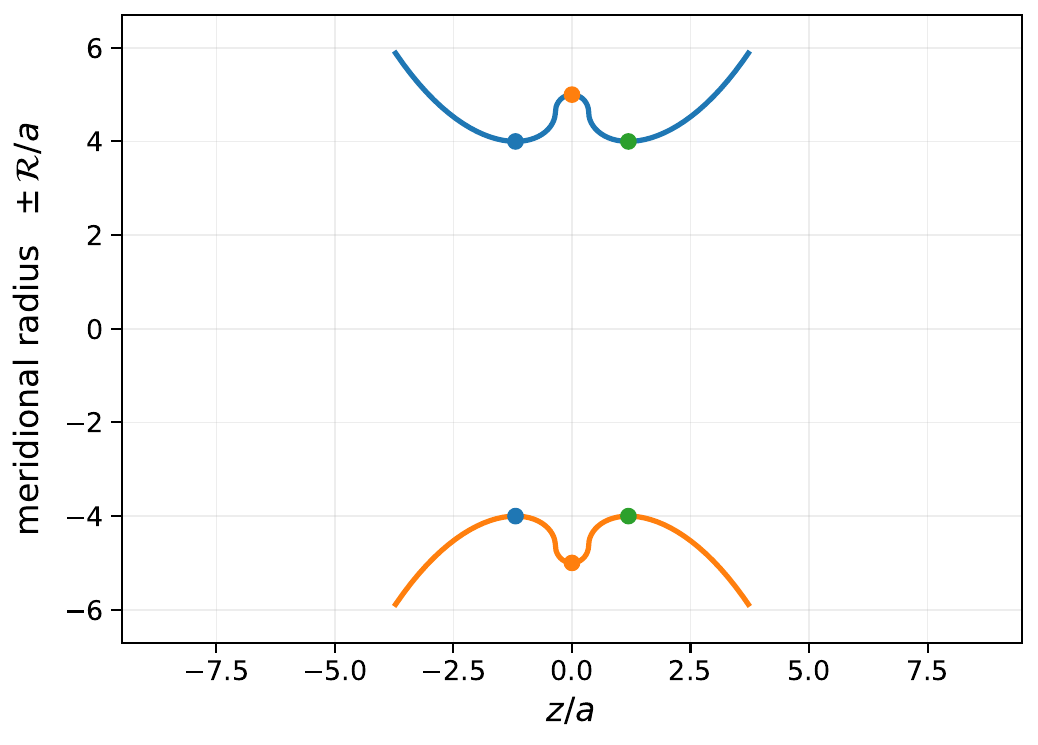}
 \caption{Exact Euclidean meridional embedding for the $N=1$, $\alpha=3$
 member.  The two narrow necks are the throats at
 $\ell/a=\pm\sqrt3$ and the enlarged central section is the anti-throat at
 $\ell=0$. The monotonic realization is $C^1$, with the endpoint regularity discussed in the text.}
 \label{fig:N1embeddingexact}
\end{figure}

At $\alpha=3$ the radicand is a perfect square,
\begin{equation}
 1-\mathcal R_1'^2
 =\frac{\left[1-3(\ell/a)^2\right]^2}
 {\left[1+(\ell/a)^2\right]^3},
 \label{eq:N1embedalpha3}
\end{equation}
and a monotonic embedding coordinate is
\begin{equation}
 \frac{z(\ell)}{a}
 =\int_0^{\ell/a}
 \frac{|1-3u^2|}{(1+u^2)^{3/2}}du.
 \label{eq:N1zalpha3}
\end{equation}
The corresponding meridional section is shown in
Fig.~\ref{fig:N1embeddingexact}.

\section{Multi-cavity null dynamics in the hierarchy}
\label{sec:multicavity}

The general optical theorem of Sec.~\ref{subsec:opticaltheorem} becomes
especially transparent in the factorized family.  Each of the $N+1$ throats
is an unstable circular null orbit and each of the $N$ anti-throats is a
stable circular null orbit.  The null potential therefore consists of
$N+1$ barriers separated by $N$ wells.  Each finite island between
neighboring throats is an optical cavity centered on an anti-throat.

For a stationary sphere at $\ell=\ell_*$, the magnitude of the circular-orbit
impact parameter is
\begin{equation}
 u_*\equiv\frac{|L|}{E}=\mathcal R_N(\ell_*).
 \label{eq:hierarchyimpact}
\end{equation}
Here $u$ is distinguished from the Morris--Thorne shape function
$b_{\rm sh}$.  At an unstable throat orbit, $u_{\rm T}$ is a critical
scattering value for rays that can reach that barrier from the relevant
asymptotic region.  At a stable anti-throat, $u_{\rm A}$ labels the circular
orbit but is not, by itself, a critical impact parameter for scattering from
infinity; access to an inner well is controlled by the intervening throat
barriers.
The signed radial optical curvature follows from
Eqs.~\eqref{eq:hierarchysecondcenter}--\eqref{eq:hierarchysecondroots} and
\eqref{eq:localopticalfrequency}.  For example, at a nonzero stationary
sphere,
\begin{equation}
 \frac{\mathcal R_N''}{\mathcal R_N}
 \bigg|_{\ell/a=\pm\sqrt{\alpha_k}}
 =\frac{2\alpha_k
 \displaystyle\prod_{j\neq k}(\alpha_k-\alpha_j)}
 {a^2(1+\alpha_k)^{N+1/2}
 \left[\mathcal R_N(\pm a\sqrt{\alpha_k})/a\right]}.
 \label{eq:hierarchyopticalcurvature}
\end{equation}
Thus both the sign and magnitude of the local instability or trapping
frequency are analytic once the $\alpha_j$ and $C_N$ are fixed.

\begin{figure}[t]
 \centering
 \includegraphics[width=\columnwidth]{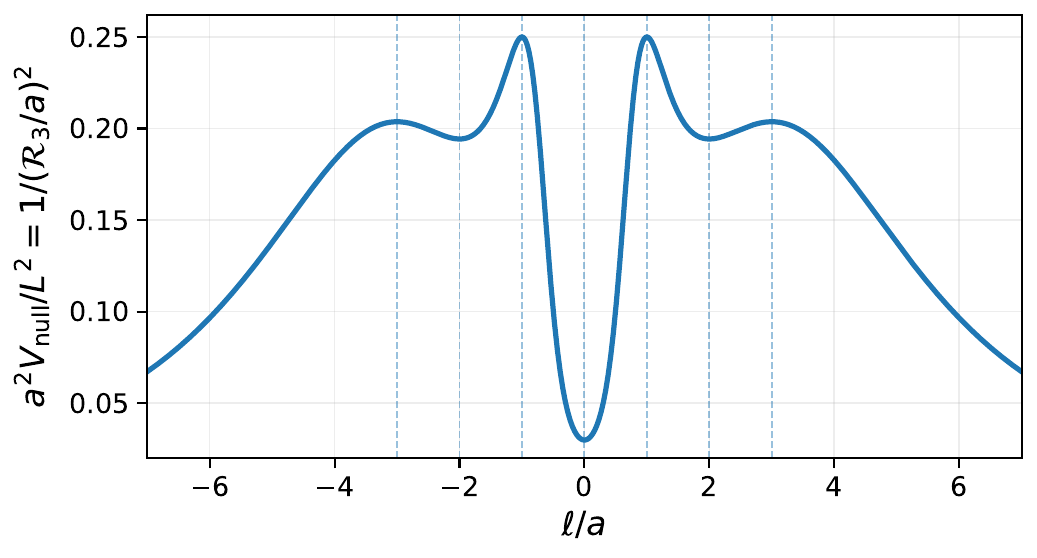}
 \caption{Dimensionless null effective potential for the explicit $N=3$
 hierarchy member with $\alpha_1=1$, $\alpha_2=4$, and $\alpha_3=9$.
 Vertical dashed lines mark the seven stationary area spheres.  The four
 maxima at $\ell/a=\pm1,\pm3$ are unstable throat light rings; the three
 minima at $\ell/a=0,\pm2$ are stable anti-throat light rings.  The geometry
 therefore contains three finite optical cavities bounded by four throat
 barriers.}
 \label{fig:N3potential}
\end{figure}

Figure~\ref{fig:N3potential} shows the dimensionless null potential for the
explicit $N=3$ solution of Eq.~\eqref{eq:N3explicit}.  The integration
constant is chosen only for visualization so that the smallest throat radius
is $2a$.  The maxima at $\ell/a=\pm1$ and $\pm3$ are the four unstable
throat light rings.  The minima at $\ell/a=0$ and $\pm2$ are the three
stable anti-throat light rings.  The result is an exact three-cavity optical
chain.

The cellwise result \eqref{eq:cellanecnegative} applies separately to each
interval bounded by adjacent stationary spheres.  For the $N=3$ example,
all six finite stationary-sphere cells have negative radial null-energy
integrals, even though the three anti-throats themselves satisfy the full set
of pointwise energy conditions.  The two exterior intervals are negative as
well.  Grouping two adjacent finite cells around each anti-throat shows that
each complete island bounded by two throats also carries a negative
integrated radial null energy.

Stable light rings support long-lived trapping and slow decay and can provide
a channel for nonlinear or secular behavior in horizonless compact
geometries \cite{Cardoso2014,Keir2016,Cunha2017}.  Their presence alone is
not, however, a universal instability theorem; fully nonlinear evolutions
have exhibited long-term stability in at least one theory-dependent setting
\cite{Guo2025}.  In the present chain, the physically relevant question is
therefore how trapped wave packets leak through, and couple across, the
successive throat barriers.  Multi-barrier and multi-well potentials motivate
searches for tunneling splittings, resonant transmission, echoes, and
long-lived modes.  Resonances have already been found in wormhole scattering
studies \cite{Pradhan2026,Furuta2026}, while multiple photon and antiphoton
spheres generate a richer lensing structure than a single barrier
\cite{Shaikh2019}.  The hierarchy makes the number and location of all such
orbits analytic by construction, providing a controlled background for
spectral and lensing calculations.

\section{Traversability, interpretation, and physical scope}
\label{sec:scope}

All hierarchy members are ultrastatic and horizonless.  Static observers are
geodesic, radial null curves satisfy
$ dt=\pm d\ell $,
and there is no gravitational redshift between the two asymptotic ends.  In
the static orthonormal frame the mixed time-space Riemann components vanish.
The relevant spatial radial-angular curvature is
\begin{equation}
 R_{\hat\ell\hat\theta\hat\ell\hat\theta}
 =-\frac{\mathcal R_N''}{\mathcal R_N}.
 \label{eq:hierarchytidalstatic}
\end{equation}
A traveler moving radially with speed $v$ and Lorentz factor
$\gamma=(1-v^2)^{-1/2}$ relative to the static frame measures, in the
traveler's primed orthonormal frame, the lateral tidal component
\begin{equation}
 \left|
 R_{\hat0'\hat\theta'\hat0'\hat\theta'}
 \right|
 =\gamma^2v^2
 \left|\frac{\mathcal R_N''}{\mathcal R_N}\right|.
 \label{eq:hierarchytidalboost}
\end{equation}
At fixed dimensionless parameters and fixed $C_N$, all curvature scales vary
as $a^{-2}$.  Thus one may enlarge the entire geometry without changing its
dimensionless throat pattern and make tidal accelerations arbitrarily small,
in the same scaling sense used in the original Morris--Thorne traversability
analysis \cite{MorrisThorne1988}.

The interpretation of the anti-throat regions should remain precise.  A
finite island is an enlarged portion of a connected spatial slice; it is not
by itself a new asymptotic universe.  The constant-time topology remains
$\mathbb R\times S^2$ for every member of the hierarchy.  The phrase
``baby-universe-like'' refers only to the local morphology of an expanded
interior chamber screened by narrower necks.  Genuine baby-universe creation
involves a dynamical causal and topological history of a different kind
\cite{Blau1987,FarhiGuthGuven1990}.

The construction separates geometrical statements from a choice of matter realization. The Einstein tensor defines an exact conserved anisotropic source, and Appendix~\ref{app:source} gives one explicit realization through a scalar field with a variable kinetic coefficient and a scalar-dependent Maxwell coupling. This reconstruction does not make the source unique or guarantee healthy perturbative degrees of freedom: the kinetic coefficients can vanish or change sign. The geometrical and optical results above depend only on the total stress tensor. Explicit scalar--NED reconstructions also support multi-throat geometries through concrete field sectors~\cite{Crispim2025}, while higher-curvature theories can
shift part of the required effective NEC violation into the gravitational
sector \cite{Antoniou2020,Ibadov2020}. Appendix~\ref{app:source} therefore provides background field support for the present hierarchy, while the viability and perturbations of that support remain separate questions.

Static existence also does not imply dynamical stability.  The
Ellis--Bronnikov wormhole is unstable under spherical perturbations
\cite{Gonzalez2009I,Gonzalez2009II}, and matter-filled double-throat systems
may retain unstable modes \cite{Dzhunushaliev2014NS,Dzhunushaliev2014Boson}.
The stable anti-throat light rings add a separate trapping channel whose
dynamical role should be examined by perturbing the complete matter--gravity
system.  The arbitrary-$N$ family is particularly suitable for asking how
instability timescales and trapped-mode spectra scale with the number,
separation, and size of the cavities.

The background stress tensor alone does not uniquely determine a coupled perturbation problem. Appendix~\ref{app:source} supplies one possible action-level realization, but its dynamical stability has not been analyzed. Such an analysis must perturb the chosen scalar, electromagnetic field, and metric with the reconstructed functions held fixed, and address any zeros or sign changes of the kinetic coefficients. The geodesic quantities
$\Lambda_{\rm T}$ and $\omega_{\rm A}$ diagnose only the radial behavior of
test null rays; they are not eigenvalues of the full matter--metric system.

Finally, every member of the factorized hierarchy has zero ADM mass because
its radial asymptotics satisfy
$1-\mathcal R_N'^2=O(\mathcal R_N^{-2})$, rather than containing the
$2M/\mathcal R_N$ term associated with a nonzero mass $M$.  Ultrastaticity
alone does not enforce this falloff.  The zero-mass property is analytically
useful but limits direct astrophysical interpretation.  A major extension
would preserve the factorized stationary-sphere pattern while introducing a
mass term in the radial asymptotics and, independently, a nontrivial redshift
function.  The latter would model gravitational redshift, whereas the former
would generate finite ADM mass.  The finite-mass two-throat construction of
Ref.~\cite{Pradhan2026} shows that this goal is feasible at the first
nontrivial multiplicity.

\section{Discussion and conclusions}
\label{sec:conclusions}

The purpose of this work was to replace isolated multi-throat examples by an
exact construction in which the entire radial architecture is controlled
analytically.  In the global proper-distance gauge, this is achieved by
prescribing
\begin{equation*}
 \mathcal R_N'(\ell)
 =\frac{(\ell/a)
 \prod_{j=1}^{N}[(\ell/a)^2-\alpha_j]}
 {[1+(\ell/a)^2]^{N+1/2}},
\end{equation*}
where the ordered positive parameters $\alpha_j$ locate the noncentral
stationary spheres.  The form of this prescription is dictated by the
problem: its numerator inserts an arbitrary finite set of symmetric simple
zeros, while its positive denominator supplies the matching asymptotic
degree, introduces no additional stationary points, and preserves elementary
integrability.  The integration constant $C_N$ then changes the absolute
areal scale without moving any extremum.  It consequently changes physical
radii, stresses, and optical frequencies even though the proper-distance
locations and classifications remain fixed.  Positions, classifications,
and sizes are thus separated into transparent algebraic data rather than
being physically independent.

The construction yields exactly $N+1$ throats and $N$ anti-throats.  This
count is not merely a feature of the chosen rational family.  It realizes a
general property of every smooth, two-ended, asymptotically flat area profile
with finitely many nondegenerate stationary spheres: the asymptotic signs of
$\mathcal R'$ force the first and last extrema to be minima and all
intermediate extrema to alternate.  A multi-throat wormhole should therefore
be viewed as a chain of finite islands, each bounded by two necks and centered
on an anti-throat.  The explicit $N=2$ and $N=3$ members show how this picture
extends from one central chamber to two and three coupled optical cavities
without changing the global topology.

The most significant physical statements are independent of the hierarchy.
In four-dimensional ultrastatic spherical Einstein gravity,
\begin{equation*}
 \rho+p_\ell=-\frac{\mathcal R''}{4\pi\mathcal R}.
\end{equation*}
The flare-out condition therefore enforces radial NEC violation at every
nondegenerate throat.  At an anti-throat the sign reverses, and the remaining
Einstein equations ensure that the NEC, WEC, DEC, and SEC all hold locally,
with $\rho+p_\ell+2p_t=0$ saturated.  The enlarged center of a cavity is thus
locally nonexotic as a consequence of ultrastatic geometry itself.  What the
additional necks accomplish is not the removal of exotic matter, but its
spatial redistribution away from the cavity centers and toward the
flare-out regions.

This distinction between local and averaged behavior is made precise by
\begin{equation*}
 \int(\rho+p_\ell)d\ell
 =-\frac{1}{4\pi}
 \int\left(\frac{\mathcal R'}{\mathcal R}\right)^2d\ell<0.
\end{equation*}
This reduced form holds whenever the boundary term
$[\mathcal R'/\mathcal R]$ vanishes: on every finite interval bounded by
consecutive stationary spheres, on each exterior interval, and on the
complete two-ended geodesic.
Consequently, every cell has a strictly negative affine radial null-energy
integral and their sum gives a strictly negative complete two-ended ANEC.
Pointwise energy conditions may all be restored at an anti-throat, but no
geometrical cell can become nonexotic in the corresponding averaged sense.
This local--global separation is one of the most robust conclusions of the
analysis.

The same radial geometry organizes null propagation.  Every stationary
sphere supports a circular null orbit, and
\begin{equation*}
 V_{\rm null}''\big|_*
 =\frac{8\pi L^2}{\mathcal R_*^2}(\rho+p_\ell)_*.
\end{equation*}
Throat light rings are therefore radially unstable, whereas anti-throat light
rings are radially stable.  An order-$N$ member contains $N+1$ optical
barriers and $N$ trapping wells.  The matter distribution and the optical
structure are not separate properties: the same local stress-energy
combination fixes the flare-out sign, the type of stationary sphere, and the
curvature of the null potential.  This identity provides a direct bridge from
energy-condition diagnostics to observable propagation phenomena.

The $N=1$ member illustrates these results without obscuring them by
algebraic complexity.  It connects the Ellis--Bronnikov geometry at
$\alpha=-1$ to a quartically degenerate critical throat at $\alpha=0$ and to
a double-throat geometry for $\alpha>0$.  Its exact source, affine
null average, proper-volume measure, island volume, branchwise
Morris--Thorne description, and Euclidean embedding criterion demonstrate
that the transition from one neck to two is simultaneously geometrical,
energetic, and optical.  At the same time, the arbitrary-$N$ construction
shows that this transition is the first step in a general sequence rather
than a special isolated phenomenon.

Several limitations delimit the interpretation.  Ultrastaticity removes
gravitational redshift, whereas the stronger radial falloff chosen for this
particular hierarchy yields zero ADM mass.  Neither statement should be
conflated with the other.
The Einstein tensor supplies a conserved effective anisotropic source, and Appendix~\ref{app:source} realizes it with explicit scalar and electromagnetic couplings. This is one background reconstruction, not a unique microscopic model or a proof of perturbative viability. Moreover, geodesic
stability of an anti-throat light ring does not imply dynamical stability of
the full matter--gravity system. The coupled perturbations of the reconstructed action remain to be studied; stable trapping may generate very slow decay and
long-lived perturbative channels, but it is not by itself a universal proof
of nonlinear instability.  The present solutions should therefore be
understood as analytic backgrounds on which these questions can be separated
and tested, rather than as complete astrophysical models.

These limitations point directly to future work.  A first bridge is to
generalize the radial asymptotics so that a $2M/\mathcal R$ mass term is
present, and independently to introduce a nontrivial redshift function while
preserving the factorized stationary-sphere data.  This would test which
energy--optical identities survive beyond ultrastaticity and generate
finite-mass chains with controlled horizons or horizonless limits.  A second bridge is to develop the reconstruction in Appendix~\ref{app:source}: one may ask whether selected subfamilies belong to a single fixed scalar--electromagnetic action, whether alternative NED or higher-curvature realizations improve their physical viability, and how the required effective NEC violation is redistributed among sectors.  A third bridge is spectral. The analytic multi-well potential is well suited to computing quasinormal modes,
tunneling splittings, echoes, transmission resonances, and branch-resolved
lensing as functions of the number and separation of the cavities.  Finally,
perturbing a specified complete source and metric will determine whether
increasing multiplicity merely reproduces the known Ellis-sector instability
or produces new collective modes and parametrically separated timescales.
Because all stationary locations remain algebraic input, these extensions
can retain the global control that motivated the construction in the first
place.

\acknowledgments{
FSNL acknowledges support from the Funda\c{c}\~{a}o para a Ci\^{e}ncia e a Tecnologia (FCT) Scientific Employment Stimulus contract with reference CEECINST/00032/2018, and funding through the research grant UID/04434/2025.
MER thanks Conselho Nacional de Desenvolvimento Cient\'ifico e Tecnol\'ogico - CNPq, Brazil, for partial financial support.
GIR acknowledges support from the FCT through the Fellowship 2026.04495.BD.
}

\appendix

\section{Einstein--Maxwell--scalar source}
\label{app:source}

The effective source derived in Sec.~\ref{subsec:energytheorem} and evaluated explicitly for $N=1$ in Sec.~\ref{subsec:N1source} can be realized by coupled scalar and electromagnetic fields. We reconstruct that source for the hierarchy of Sec.~\ref{sec:hierarchy}, using the same metric~\eqref{eq:generalmetric} with $\mathcal R=\mathcal R_N$. The scalar $\varphi$ has a kinetic coefficient $\epsilon(\varphi)$, potential $V(\varphi)$, and a scalar-dependent electromagnetic coupling $W(\varphi)$. A representative action is
\begin{equation}
  S=\int d^4x\sqrt{-g}\left[\frac{1}{2\kappa^2}R
   -\epsilon(\varphi)(\partial\varphi)^2
   -V(\varphi)-W(\varphi)F\right],
  \label{eq:appaction}
\end{equation}
where $\kappa^2=8\pi$ and $F=\frac14F_{\mu\nu}F^{\mu\nu}$ is the Maxwell invariant. Here $R$ is the spacetime Ricci scalar and $(\partial\varphi)^2\equiv g^{\mu\nu}\partial_\mu\varphi\partial_\nu\varphi$. A positive $\epsilon$ corresponds to a canonical scalar field and a negative one to a phantom field.

We consider a purely magnetic field with nonzero magnetic charge $q_m$, whose only nonvanishing components are $F_{\theta\phi}=-F_{\phi\theta}=q_m\sin\theta$, so that $F=q_m^2/(2\mathcal R_N^4)$. For this configuration, the Maxwell equations, $\nabla_\mu(WF^{\mu\nu})=0$, hold for any $W(\varphi)$, and the Einstein equations give
\begin{align}
  \epsilon(\ell) &= -\frac{\mathcal R_N''}{\kappa^2\,\mathcal R_N\,\varphi'^2}\,,
  \label{eq:appeps}\\[4pt]
  V(\ell) &= \frac{1-\mathcal R_N'^2-\mathcal R_N\mathcal R_N''}{2\kappa^2\mathcal R_N^2}\,,
  \label{eq:appV}\\[4pt]
  W(\ell) &= \frac{\mathcal R_N^2\left(1-\mathcal R_N'^2-\mathcal R_N\mathcal R_N''\right)}{\kappa^2 q_m^2}
     = \frac{2\mathcal R_N^4}{q_m^2}\,V(\ell)\,.
  \label{eq:appW}
\end{align}
For a monotonic scalar profile with $\varphi^\prime\ne0$ at every finite $\ell$, the scalar field equation follows from the Einstein equations, the Maxwell equation, and the contracted Bianchi identity. The profile chosen below satisfies this condition. From Eq.~\eqref{eq:appeps}, the scalar field is phantom at every throat ($\mathcal R_N''>0$) and canonical at every anti-throat ($\mathcal R_N''<0$). Comparing with Eqs.~\eqref{eq:generalrho}--\eqref{eq:generalpt}, one finds $\epsilon\,\varphi'^2=(\rho+p_\ell)/2$ and $V=(\rho+p_t)/2$.

For the scalar field we choose the Ellis--Bronnikov profile~\cite{Ellis1973,Bronnikov1973},
\begin{equation}
  \varphi(\ell)=\frac{1}{\kappa}\arctan\!\left(\frac{\ell}{a}\right),
  \label{eq:appphi}
\end{equation}
for which, when $N=0$ and $C_0=0$, Eqs.~\eqref{eq:appeps}--\eqref{eq:appW} give $\epsilon=-1$ and $V=W=0$, reproducing the free phantom scalar field that supports the Ellis--Bronnikov wormhole. Since $\varphi$ is monotonic, it can be inverted as
\begin{equation}
  \ell=a\tan(\kappa\varphi)\,,\quad
  \kappa\varphi\in\left(-\tfrac{\pi}{2},\tfrac{\pi}{2}\right),
  \label{eq:appphiinv}
\end{equation}
which allows $\epsilon$, $V$ and $W$ to be written explicitly as functions of $\varphi$.

In the particular case $N=1$ with $C_1=0$, the areal radius is given by Eq.~\eqref{eq:N1profile}, which in terms of the scalar field reads $\mathcal R_1(\varphi)=a\,[1+(1+\alpha)\cos^2(\kappa\varphi)]/\cos(\kappa\varphi)$. Substituting into Eqs.~\eqref{eq:appeps}--\eqref{eq:appW} then gives
\begin{align}
  \epsilon(\varphi) &= \frac{3(1+\alpha)\cos^2(\kappa\varphi)-(3+2\alpha)}
                            {1+(1+\alpha)\cos^2(\kappa\varphi)}\,,
  \label{eq:appepsN1}\\[6pt]
  V(\varphi) &= \frac{(1+\alpha)^2\cos^6(\kappa\varphi)\left[4\cos^2(\kappa\varphi)-3\right]}
                     {2a^2\kappa^2\left[1+(1+\alpha)\cos^2(\kappa\varphi)\right]^2}\,,
  \label{eq:appVN1}\\[6pt]
  W(\varphi) &= \frac{a^2(1+\alpha)^2}{\kappa^2 q_m^2}\,
                \cos^2(\kappa\varphi)\left[4\cos^2(\kappa\varphi)-3\right] \nonumber\\
             &\quad\times\left[1+(1+\alpha)\cos^2(\kappa\varphi)\right]^2 .
  \label{eq:appWN1}
\end{align}
The same procedure gives closed-form expressions for any $N$. These functions reproduce the total stresses of Sec.~\ref{subsec:N1source}; the phantom/canonical change in $\epsilon$ follows the same $\mathcal R_N^{\prime\prime}$ sign that controls the radial NEC and optical classification in Sec.~\ref{sec:universal}. The field-dependent coefficients, including $W$, need not remain positive or nonzero, so background reconstruction does not establish a healthy or well-posed perturbation theory. In general the reconstructed functions depend on the hierarchy parameters; varying those parameters therefore need not describe different solutions of one fixed matter action. This is the limitation invoked in Sec.~\ref{sec:scope}.

\onecolumngrid
\section{Comparison with representative multi-throat models}
\label{app:comparison}

Table~\ref{tab:comparison} complements the literature discussion in Sec.~\ref{sec:intro} by separating the geometrical strategies, asymptotic properties, and scope of representative recent constructions. The comparison identifies complementary approaches rather than ranking them or assigning priority to features already established elsewhere.

\begin{table}[!ht]
\hypersetup{citecolor=blue,linkcolor=blue}
	\centering
	\small
	\setlength{\tabcolsep}{5pt}
	\renewcommand{\arraystretch}{1.15}
	
	\caption{Relation of the present construction to representative recent
		multi-throat models. The table is intended to delimit scope rather than
		to rank the approaches.}
	\label{tab:comparison}
	
	\begin{tabular}{
			@{}
			>{\raggedright\arraybackslash}p{0.15\textwidth}
			>{\raggedright\arraybackslash}p{0.23\textwidth}
			>{\raggedright\arraybackslash}p{0.19\textwidth}
			>{\raggedright\arraybackslash}p{0.33\textwidth}
			@{}
		}
		\toprule
		\textbf{Reference}
		& \textbf{Geometrical strategy}
		& \textbf{Asymptotic/source feature}
		& \textbf{Main distinction relative to the present work}
		\\
		\midrule
		
		Crispim \emph{et al.} \cite{Crispim2025}
		& Prescribed multi-extremum areal function; scalar--NED reconstruction
		& Asymptotically flat; explicit field sources
		& Multiple throats/anti-throats and multiple stable/unstable null orbits
		are already present, but the stationary locations are not generated
		by an arbitrary factorized $N$-parameter hierarchy with a closed
		counting theorem.
		\\
		\midrule
		
		Cataldo \emph{et al.} \cite{Cataldo2026}
		& Quartic embedding ansatz producing exactly two throats
		& Asymptotically flat; zero ADM mass
		& Provides a systematic two-throat embedding construction; the present
		method works directly in global proper distance and extends
		analytically to arbitrary finite multiplicity.
		\\
		\midrule
		
		Amaral \emph{et al.} \cite{Amaral2026}
		& Localized Ellis--Bronnikov deformation
		& String-cloud/global-monopole-type asymptotics
		& Demonstrates a nonexotic central belly in a specific double-throat
		model; here the nonexoticity of every ultrastatic anti-throat is
		proved as a general theorem.
		\\
		\midrule
		
		Pradhan \emph{et al.} \cite{Pradhan2026}
		& Finite-mass deformation of a two-throat embedding profile
		& Standard asymptotic flatness with finite mass
		& Develops global light rings, lensing, and wave transmission for two
		throats; here the emphasis is arbitrary multiplicity and universal
		local/averaged identities.
		\\
		\midrule
		
		Present work
		& Factorized $\mathcal R_N'(\ell)$ with arbitrary ordered roots
		& Asymptotically flat; ultrastatic; zero ADM mass from radial falloff
		& Exact $N+1$ throat/$N$ anti-throat hierarchy, universal
		energy--optical correspondence, and negative radial null-energy
		integral in every stationary-sphere cell.
		\\
		\bottomrule
	\end{tabular}
\end{table}

\twocolumngrid

\end{document}